\documentclass[%
 reprint,
nofootinbib,
nobibnotes,
 amsmath,amssymb,
 aps,
pra,
]{revtex4-2}

\usepackage{amssymb, amsmath}

\usepackage[dvipsnames]{xcolor}

\usepackage{ulem, enumitem} 
\usepackage{graphicx}
\usepackage{hyperref}

\newcommand{\eqn}[1]{\begin{eqnarray} #1 \end{eqnarray}}
\newcommand{\tit}[1]{\textit{#1}}
\newcommand{\tbf}[1]{\textbf{#1}}
\newcommand{\trm}[1]{\textrm{#1}}

\newcommand{\trc}[2]{  \textrm{tr}_{#1}\left[ #2 \right]  }
\newcommand{\zum}[2]{\displaystyle\sum_{#1}^{#2}}

\newcommand{\ket}[1]{| #1 \rangle}
\newcommand{\ketbra}[2]{| #1 \rangle \langle #2 |}
\newcommand{\braket}[2]{\langle #1 |  #2 \rangle}

\begin{document}

\preprint{APS/123-QED}

\title{What does quantum parameter estimation estimate? \\ A no-go result for true quantity values}

\author{Jacques L Pienaar}
\email{jacques.pienaar@zoho.com}
\affiliation{Instituto de Física, Universidade Federal do Rio de Janeiro, Caixa Postal 68528, Rio de Janeiro, 21941-972, RJ, Brazil
}

\author{Eran Tal}
\author{Asya Ciftci}
\affiliation{Department of Philosophy, McGill University, 855 Sherbrooke St W, Montreal, H3A 2T7, Quebec, Canada}

\date{\today}

\begin{abstract}
We examine the conditions under which quantum parameter estimation is compatible with the concept of a ``true'' quantity value: a unique value of the estimated quantity that exists prior to and independently of measurement. While quantum parameter estimation typically assumes that the quantity to be estimated is a parameter, we argue that this assumption can be lifted, opening the way to a more general formulation of estimation that may apply even in situations where the quantity to be estimated is no longer a parameter, but instead a quantum variable susceptible to quantum fluctuations. We identify three assumptions that characterize the way estimation is often presented in the quantum metrology literature, namely, as a procedure for inferring an unknown ``true value", which is assumed to be unique and to exist prior to and independently of measurement. Using our more general framework, we prove a no-go theorem showing that the preceding assumptions are jointly incompatible with two further assumptions: the completeness of the quantum state (no `hidden variables'), and the presence of entanglement between the probe systems and the system under measurement. We note that the presence of such entanglement is a generic prediction of standard quantum theory. Consequently, if one is not willing to accept hidden variables, nor willing to modify standard quantum theory by introducing a new physical mechanism that destroys entanglement at some scale or regime, then a realist account of estimation is ultimately impossible.
\end{abstract}

\maketitle

\section{Introduction}
\label{intro}

Quantum parameter estimation is not, despite its name, normally the estimation of a quantum parameter. It is the use of quantum systems, quantum dynamics, and quantum measurements to estimate a parameter that is assumed to be classical. The parameter may control a quantum state, a Hamiltonian, a channel, or a measurement; but the parameter itself is an ordinary number, drawn from some set of alternatives, and the task of estimation is to discover which number it is. This much is explicit in standard presentations of quantum estimation theory, from Helstrom and Holevo to more recent treatments of quantum metrology~\cite{helstrom1976quantum,holevo2010probabilistic,giovannetti_advances_2011,paris_quantum_2009,escher_general_2011,montenegro_review,polino_2020}.

By a \tit{parameter}, we shall mean a quantity represented in the model by an element $\theta$ of a set $\Theta$ (discrete or continuous), such that exactly one value of $\theta$ is taken to obtain in each run of the experiment, whether or not it is known to the experimenter. By a \tit{quantum variable}, by contrast, we shall mean a physical quantity associated to an observable, here denoted $\hat{\theta}$, whose spectrum of eigenvalues is $\Theta$, defined on a quantum system whose state need not be an eigenstate of that observable, and whose relevant degrees of freedom may become entangled with other systems. (These definitions will be made more precise in Sec.~\ref{sec:parameters}). This distinction is deliberately modest: it does not aim to designate two kinds of things in nature, but between two kinds of representation in a physical model. 

The distinction nevertheless matters. Estimation, as the word is ordinarily used, aims to reveal the value of something that is not already known. If the quantity of interest is modeled by a parameter, this is straightforward: the parameter is modeled as having a value prior to and independently of its estimation; the experimenter just doesn't know what the value is. To the extent that the model is taken to describe reality, this moreover supports a realist view of the parameter value as corresponding to a quantity that exists already, before measurement reveals it. 

If, on the other hand, the quantity of interest is modeled by a quantum variable, the same statement is no longer straightforward. A quantum state need not assign any determinate value to $\hat{\theta}$ prior to measurement; in particular it may assign a superposition of different $\hat{\theta}$ eigenstates. To speak of estimating the value of such a variable therefore risks incoherence. It is not obvious what the value is supposed to be, nor when it is supposed to have become available for estimation. Historically, attempts to postulate that measuring $\hat{\theta}$ reveals a pre-existing value --- known in the literature as a \tit{hidden variable} --- have encountered serious challenges, most notably Bell's Theorem; we will review this history in Sec.~\ref{sec:Bell}. 

The fact that many physical quantities that are conventionally modeled as parameters might equally well be modeled as quantum variables is easy to miss because many quantities of practical interest behave, for all experimental purposes, as parameters. A bath temperature, a macroscopic displacement, or a magnetic field produced by a stable macroscopic current can often be represented by a number whose possible quantum fluctuations are irrelevant. In such cases, the usual framework is justified, at least pragmatically. This makes it all too easy to conflate the decision to model a quantity as a parameter, with a claim about the nature of the physical quantity in itself, \textit{viz.}\ that it \textit{is} a parameter. If one falls into this trap, then one may be led to make inconsistent predictions when the same quantity appears in a new experimental context that amplifies its hitherto unnoticed quantum aspects.

Consider magnetic field estimation. A spin-$1/2$ atom may be used to probe the magnetic field component $F$, with the field producing a rotation of the spin by an angle $\eta(F)$~\cite{zhang_molmer_2020,apellaniz_2018,troiani_2018}. If the field is generated by a large current in a warm, noisy, uncontrolled conductor, it may be entirely appropriate to treat $F$ as a parameter. In this case, each probe leaves the interaction in a pure quantum state $\ket{\psi_{\eta(F)}}$ that depends on $F$; knowing the state of the probes thus entails perfect knowledge of the angle $\eta$ which can be inverted to obtain $F$, so the problem reduces to that of estimating an unknown quantum state given $N$ copies of it. By contrast, if the field is generated by a carefully isolated superconducting circuit whose current can be put into a superposition of macroscopically distinct directions, then the corresponding magnetic field component may not be classical in the relevant sense. In this case, each probe becomes entangled to the field, and emerges in a state corresponding to a statistical mixture of different possible angles $\eta(F)$. Moreover, since the joint state of field plus probes is an entangled superposition of different values of $F$ and associated angles of the probes, one cannot just presume that the estimation obtained from measuring the probes' angle corresponds to a determinate value of $F$ that existed prior to performing this measurement --- this would be tantamount to postulating a hidden variable for the quantized field magnitude. Ultimately, what settles the issue is not the word ``magnetic'', but the physical context: the degree of decoherence, the accessibility of the source degrees of freedom, and the possibility of detecting interference or entanglement involving the field.

The interesting case is the intermediate one. Suppose we wish to estimate a field produced by a mesoscopic current source, engineered well enough that coherent current superpositions are not obviously impossible, but not well enough that decoherence can be assumed to have selected a unique classical value prior to the probe interaction. Or suppose that a phase shift in an interferometer is produced by a small object whose internal state is not fully controlled, so that the same phase-estimation protocol might be probing either a classical optical element or a quantum object capable of becoming entangled with the light. Such examples are artificial, but not fantastical. They ask a simple methodological question: what should we do when we are uncertain whether the quantity of interest is a parameter or a quantum variable?

The obvious answer is to use the more general model. Quantum mechanics contains classical mechanics as a limiting case, or at least as an effective description obtained under suitable restrictions such as decoherence, coarse-graining, or superselection. If there is doubt about the status of the quantity of interest, one should model it quantum mechanically and recover the classical case as a special case. This is the strategy we pursue in this article.

The strategy immediately encounters a difficulty. Once the quantity of interest is modeled as a quantum variable, it is no longer clear that we are doing parameter estimation at all. Can an estimation protocol be defined if the value to be estimated is not assumed to be a classical number before the protocol begins? If a measurement of $\hat{\theta}$ would yield an eigenvalue, what entitles us to identify that eigenvalue with a pre-existing value of the quantity? And if we do make such an identification, have we not reintroduced precisely the sort of hidden-variable commitment that Bell-type no-go theorems taught us to treat with care~\cite{einstein1935can,zukowski_2014,kooderi_suresh_thesis,bell_tests,sep-bell-theorem}?

Our aim is to answer these questions without prejudging the interpretation of quantum theory. We proceed in three steps. 

First, we show how standard quantum parameter estimation can be embedded in a model in which the quantity of interest is itself represented quantum mechanically. In this model, the usual parameter is encoded as the eigenvalue of an associated quantum observable $\hat{\theta}$, which serves to represent a hypothetical ideal measurement of the parameter value; nevertheless, in the special case where the quantum state is diagonal in the basis selected by $\hat{\theta}$, one recovers precisely the same predictions as for a parameter $\theta$.

Second, we isolate three assumptions under which the resulting protocol still deserves to be called an estimation protocol --- which may apply even in the general case when the state encoding the unknown parameter is non-diagonal. These assumptions are: \tit{Uniqueness}, that exactly one hypothesis about the quantity of interest is true at the end of the protocol; \tit{Pre-existing Value}, that the hypothesis singled out at the conclusion of estimation was already true before the probes interacted with the quantity of interest; and \tit{Independence}, that the truth of the hypothesis is equivalent to a claim about the value that a measurement of $\hat{\theta}$ would have revealed, whether or not that measurement was actually performed. These assumptions are close relatives of the older metrological ideal of a ``true value'', as discussed in the \textit{International Vocabulary of Metrology} (VIM) and in the literature on the Error and Uncertainty approaches to measurement~\cite{VIM_1984,VIM_2012,mari2023measurement,GrégisFabien2015Cwdw,JCGM_GUM_2008,Mari_Giordani}. We do not claim that metrologists are committed to this ideal. Rather, we use these assumptions to spell out what has to be true if estimation is to be understood as revealing a value of the quantity of interest.

Third, we introduce two further assumptions that become natural once the quantity of interest is treated quantum mechanically. \tit{Entanglement} says that the imprinting interaction between the system $A$ and probes $B$ may entangle them, even when the system bearing the quantity of interest is macroscopic and highly decoherent. \tit{Completeness} says that the quantum state expresses all relevant information about the quantity of interest (hence that there are no \tit{hidden variables}). This has the formal consequence that if a measurement of $\hat{\theta}$ would reveal a unique value with certainty, the quantum state of the system to which $\hat{\theta}$ pertains must not be a quantum superposition of eigenstates of $\hat{\theta}$. We then prove that, within our generalized model of estimation, these two assumptions are jointly inconsistent with Uniqueness, Pre-existing Value, and Independence, thus excluding a minimal realist interpretation of the estimation of the quantity of interest. 

It should come as no surprise that parameter estimation fails in general when the parameter is replaced in the model by a quantum variable. Our contribution is to articulate the set of inconsistent assumptions underlying this failure. It is especially noteworthy that the five assumptions we identify are minimal, in the sense that rejecting any \tit{one} of them is sufficient to restore consistency among the remaining four. This allows us to analyze precisely how realist commitments about measurement (the first three assumptions) interlock with assumptions about reality in quantum theory (the last two assumptions). 

Our result could therefore be read in two ways:

For those committed to an interpretation of quantum theory that upholds both Entanglement and Completeness, the conclusion is that estimation of a physical quantity can only be regarded as revealing a pre-existing true value \tit{only if} one is justified in modeling the quantity by a parameter. In particular, for quantities which are believed to be fundamentally quantum mechanical in nature --- energy, momentum, field magnitudes, \tit{etc.} --- the fact that they can be modeled as parameters under certain conditions does not license us to interpret the estimation process as providing information about a true value that existed prior to and independently of measurement. Interestingly, as we will see, this conclusion is already anticipated and embraced by \textit{coherentist} interpretations of measurement, which do not entail such a strong metaphysical reading of the estimation process.

Alternatively, if one is prepared to adjust one's interpretation of quantum theory by dropping either Entanglement or Completeness (thereby introducing either an `objective collapse' mechanism, or hidden variables, respectively) our result shows how one can --- remarkably --- construct a meaningful definition of ``estimation of a quantum variable" that retains its strong metaphysical reading as revealing a quantity value that exists prior to and independently of measurement.

Thus, between these two alternatives, our analysis does not merely prove that quantum variables cannot be estimated; we examine in detail how realist commitments about measurement in metrology intersect with realist commitments about the meaning of the quantum state. To our knowledge, this is the first time that concepts from quantum foundations have been systematically brought to bear upon the specialized domain of quantum estimation. 

The article is organized as follows. Sec.~\ref{sec:background} introduces the basic concepts that we use in this work, illustrating them with the help of an example of estimating the position of a quantum particle, and explaining how we aim to pose the central question: `can quantum variables be estimated?' Sec.~\ref{sec:framework} formalizes the estimation procedure in terms of a `controlled-circuit model' that serves as an abstract definition for both classical and quantum estimation and lays the groundwork for proving our main result. In Sec.~\ref{sec:problem_QPE} we present and prove the main result: a `no-go theorem' establishing the mutual inconsistency of a set of assumptions concerning the interpretation of estimation in a quantum setting. The discussion (Sec.~\ref{sec:discussion}) is divided into three parts: Sec.~\ref{sec:Bell} explains the relation of our result to Bell's theorem; Sec.~\ref{sec:de_finetti} shows how one can avoid inconsistency by adopting a more stringent definition of estimation that incorporates an extra `Extendibility' assumption (inspired by the de Finetti theorem); and Sec.~\ref{sec:implications} explores the implications of rejecting each of the five key assumptions used in the no-go theorem. 

\section{Background \label{sec:background}}

\subsection{Definitions: Physical quantities, parameters and quantum variables \label{sec:parameters}}

At the outset, we must be careful to define what we mean by the terms \tit{physical quantity}, \tit{parameter} and \tit{quantum variable}.  For simplicity, throughout the article we restrict ourselves to the case of estimation of a single physical quantity, which may be represented in a model by a parameter $\theta$ or, as we will show, by a quantum observable $\hat{\theta}$.

\tit{Physical quantities:} At a high level, physical quantities are the kinds of things that the physicists typically aim to measure. Examples include: energy, temperature, magnetic field magnitude, electric charge, velocity in a given reference frame, rest mass, etc. Note that this captures a quite diverse class of quantities: some are frame-dependent (velocity) while others are invariant (rest mass); some are fundamental (electric charge) while others are only defined in a suitable macroscopic limit (temperature). That being said, these examples all share the characteristics of having physical units associated to them; they take values in a real continuum; they have a metric that allows different intervals to be compared; and they represent what one could reasonably call `properties' of a physical system. When speaking of a physical quantity in this article, we mean only quantities of this kind.

On the other hand, many things that physicists measure might not have all these characteristics. The number of particles of a substance maybe regarded as a pure number without units, or alternatively may be measured in moles;  particle scattering cross-sections are probabilities, and thus it is controversial whether they constitute `properties' and the physical meaning of a probability metric is also open to debate; and quantum-mechanical phase is notoriously hard to classify due to difficulties involved in defining a phase observable. For the sake of clarity, as it is sufficient for our argument, we shall not consider these more ambiguous cases.

Before proceeding, for the sake of precision, we limit ourselves --- without losing much generality --- to systems admitting of a Hamiltonian model, where one has a notion of \tit{phase space} representing a continuous manifold upon which different states of the system are represented as functions. Classically, each point in phase space specifies the values of its generalized co-ordinates $q$ and momenta $p$, and the pair $(q,p)$ represents a possible unique configuration in which all the system's properties have precise values. Thus, any physical quantity is assumed to be determined from these by a function $A(q,p)$. These functions, and the phase-space co-ordinates themselves, will here be called \textit{variables}, and their defining characteristic is that they serve to map out the phase space, that is, the abstract space of possible `arrangements' of system properties.

By contrast, we use the term \tit{parameters} exclusively to designate numerical parameters that serve to pick out one particular Hamiltonian model from among a (typically continuous) \tit{family} of models. Unlike variables, there is no requirement that parameters represent physical properties (although they might do so); their primary function is to serve as \tit{abstract labels} that differentiate \tit{distinct hypotheses} about which model applies to the system. This interpretation makes clear the fact that all forms of parameter estimation are, at their roots, protocols for hypothesis testing --- a fact that has often been emphasized in the literature~\cite{helstrom1976quantum, jaynes_book_2003, van_trees_book, tsang_continuous_2012, kiilerich_hypothesis_2018}. This then leads to the following definitions:\\

\tit{Quantum variables:} By stipulation, a quantum variable is anything represented in the model by a quantum \tit{observable}, formally: a Hermitian (self-adjoint) operator on a separable complex Hilbert space that represents the state space of a physical system. An observable represents an ideal measurement of some physical quantity associated to the system, whose possible outcomes are given by the (discrete or continuous) eigenvalue spectrum of the operator. An important example is the Hamiltonian observable $\hat{H}$, which represents an ideal measurement of the energy of a system.\\

\tit{Parameters:} These are strictly optional features of a model, and only enter into consideration when there is some doubt on the part of the experimenter concerning the initial state or subsequent evolution. In the context of a quantum Hamiltonian model of a system, each value of a parameter $\theta$ designates a distinct hypothesis concerning (i) the state of the system, as represented by a probability function on phase space or (ii) the dynamical map, generated by the Hamiltonian function on phase space. In the former case, known as \tit{preparation estimation}, the prior distribution $P_0(\theta)$ may be regarded as expressing initial uncertainty about how the state of the system was initially prepared; in the latter case, known as \tit{channel estimation}, the prior may be seen as representing initial uncertainty about which specific dynamical map is applied to the system. \\

Let us now remark on some important aspects of these definitions.

First, every model of a system (quantum or classical) requires us to specify a set of \tit{variables} representing its properties, that is, its set of physical quantities determinable by an ideal measurement. By contrast, it is not necessary for the model to contain any \tit{parameters} of the kind we defined above. The latter only come into play when there is some uncertainty about which model applies, such as when one is explicitly ignorant about which initial state or channel is being applied. Given a model involving some parameters, one can always specify unique values for the parameters and thereby obtain a model which has no free parameters (essentially, picking out one hypothesis from among the parameterized family of hypotheses) --- indeed, this is the end goal of any estimation protocol.

Second, physical quantities can be represented in the model as parameters or quantum variables (or both --- see the example that follows): the choice of which representation to use is left to the discretion of the experimenter, based on the details of the particular experiment. 

Third, the notion of a \tit{hidden variable} in quantum mechanics --- a concept famously challenged by `no-go' theorems such as Bell's theorem --- only applies to quantum variables. The reason has to do with the very definition of a hidden variable:\\

\tbf{Definition:} Given a quantum system in a pure state $\ket{\psi}$, a \tit{hidden variable} $\lambda$ is a quantity whose value, if known, would allow one to predict the outcome of some measurement on the system with greater precision than would be implied by just the state $\ket{\psi}$ alone. More formally, if $P(k|M,\psi)$ represents the probability to obtain outcome $k$ given that one performs measurement $M$ and the state is $\ket{\psi}$, $\lambda$ is a hidden variable if there exists some choice of $k,M,\ket{\psi}$ for which $P(k|M,\psi,\lambda)$ is more informative (i.e.\ has strictly lower Shannon entropy) than $P(k|M,\psi)$. \\

Note that in quantum theory a pure state is the most accurate state possible: the formalism does not allow one to make more precise predictions than those implied by a pure state assignment. Thus, intuitively, positing the existence of a hidden variable is equivalent to asserting the \tit{incompleteness} of quantum theory, since it means there exists some information pertinent to a quantum system which is not captured by its quantum state. Whether quantum theory should be supplemented by hidden variables in order to eliminate its characteristic indeterminacy is controversial: many formal results --- most notably Bell's theorem --- have shown that the existence of hidden variables is in tension with other reasonable assumptions about how physical theories should behave.

A brief overview of the problems with hidden variables is postponed to Sec.~\ref{sec:Bell}; the salient point we wish to emphasize here is that parameters \tit{as we have defined them above} are not hidden variables. This is readily seen by noting that when one fixes the value of a parameter, one merely singles out a particular hypothesis about the \tit{quantum state} --- knowing the value of a parameter therefore cannot provide any further refinement than assigning a unique pure quantum state to a system. By contrast, knowing the value of a \tit{hidden variable} would (by definition) necessarily enable predictions strictly more precise than those implied by even a pure quantum state. 

Thus, insofar as one \tit{chooses} to model a physical quantity as a parameter, one is thereby \tit{assuming} that no inconsistency would arise if one were to further postulate that the quantity has a unique true value that exists prior to and independently of measurement; in particular one is assuming that this postulate would not be susceptible to the kinds of no-go theorems that apply to hidden variables. If it can be argued that a given quantity \tit{must} be modeled as a parameter, this need not trouble us. We are here interested in the intermediate case of quantities for which the modeling choice is \tit{ambiguous}, for in such cases the decision to model them as parameters may not be innocent. An explicit example will be given in the next section.

\subsection{Example: Estimating the mean position of a particle}
To illustrate how a single quantity can be represented as a parameter or quantum variable, consider this example from the influential textbook of Helstrom~\cite{helstrom1976quantum}: an apparatus is situated at the origin of the laboratory co-ordinate frame, and is subjected to displacement by an amount $x_0$, which may be assumed constant but unknown in every repetition of the experiment. The apparatus subsequently produces a quantum particle whose complex-valued wave-function $\psi(x-x_0)$ is a Gaussian wavepacket centered at the --- initially unknown --- apparatus position $x_0$. Subsequently, an ideal measurement is performed of the particle's position relative to the laboratory frame; this measurement may be represented by the observable $\hat{X}$. Helstrom discusses how the outcomes $(x_1,x_2,...x_N)$ obtained from measurements of $\hat{X}$ in repeated independent experiments enables an estimation of $x_0$. 

This example is highly instructive because both the outcome `$x$' of a single $\hat{X}$ measurement \tit{and} the parameter $x_0$ ostensibly provide information about one and the same physical quantity, namely, the particle's position. Despite the fact that the former is a quantum variable, while the latter is a parameter, both contribute towards answering the same question: `where can I expect to find the particle?'. More precisely, given that the apparatus is at position $x_0$, the probability for the particle to be found in an interval $[x_1,x_2]$ is given by:
\eqn{
P(x_1< x < x_2 | x_0) = \int^{x_b}_{x_a} |\psi(x-x_0)|^2 \, ,
}
which is maximum in the vicinity of the apparatus location $x_0$. Thus, the parameter $x_0$ clearly tells us something about where the particle is most likely to be found. On the other hand, the precision of the prediction is ultimately limited by the variance of the wavefunction $\psi(x-x_0)$: this represents the quantum contribution to uncertainty about where the particle will be detected and it remains nonzero even after $x_0$ is known precisely. 

Note that once $x_0$ is estimated to high precision, there is no inconsistency with assuming that the estimated value approximates the true value of the unknown displacement --- that is, one may assume that in every trial the displacement already was $x_0$ to within error bars, and the estimation procedure merely revealed that fact. 

By contrast, the interpretation of the individual outcome $x$ obtained when measuring $\hat{X}$ is not nearly so straightforward. As Helstrom points out, 
\begin{quote}
The relationship [...] of the operator $\hat{X}$ to the classical concept of the position of a particle on the x-axis tempts one to assert that the instrument measuring $\hat{X}$ is really determining the position of the particle as it was immediately before interaction with the instrument; and there is a temptation as well to set the outcome of an instrument's measurement of the momentum operator $\hat{P}$ equal to the momentum of the particle in the classical sense. If one falls into this mental trap, one easily conceives of the particle as having both a position and a momentum, and one finds fault with quantum theory for not providing a means of measuring both at the same time\footnote{Incidentally, this is an excellent concise summary of the EPR argument.}.
\end{quote}

More generally, unlike the case of estimating a parameter like $x_0$, if one similarly tries to postulate that the value revealed by measurement of a quantum variable reveals some pre-existing fact about the particle, one runs into well-known paradoxes exemplified by the EPR argument and Bell's 'no-go' Theorem. 

It is worth pausing here to reflect on why $x_0$ does not suffer from this problem. Following Helstrom, one can represent the very same wavefunction in the momentum basis, where it will again be found to be a Gaussian centered at some momentum $p_0$. It is readily seen that there is no contradiction with finding out both values $x_0$ and $p_0$ through estimation; after all, this amounts to nothing more than deciding on a fixed quantum state from among a parameterized family of possible states. Speaking loosely: the state itself still retains its \tit{quantum} uncertainty, which is what obstructs simultaneous perfect predictions of the outcomes of both $\hat{X}$ and $\hat{P}$ measurements; the estimation has merely eliminated the classical contribution to the uncertainty.

The central lesson we take from this example is that the decision to model a physical quantity as a parameter or quantum variable carries with it a great deal of hidden interpretational baggage. If one wishes to model the apparatus as a classical object --- as Helstrom chose to do in this example --- one had better provide good justification for doing so. Note since there is no known physical principle that prevents the macroscopic apparatus from being placed in a quantum superposition of positions relative to the lab co-ordinates\footnote{While decoherence is often cited as performing this function, there are ways of preserving quantum effects even at macroscopic scales, and experiments have now found quantum effects in what are arguably macroscopic systems; for a survey see Ref.~\cite{macroscopic_review}.}, it is left to the experimenter to decide whether circumstances permit modeling its unknown position as a parameter, or whether it is necessary to treat it as a quantum variable. In the latter case, one would need to then contemplate the possibility of the particle and apparatus positions becoming \tit{entangled} --- a point we will return to later. At any rate, so long as this remains even a \tit{theoretical} possibility, the realist claim that estimation reveals a pre-existing true value may be challenged.

\subsection{Can quantum variables be estimated? \label{sec:conservative_pragmatic}}

Given the definitions of the preceding sections, a question naturally arises: insofar as parameter estimation seeks to assign a value to a physical quantity, how should this quantity be represented in a model? Here we may identify two options: the \tit{conservative} view, which states that the estimand \tit{must} be modeled as a parameter, or the \tit{pragmatic} view, which states that the estimand can also be modeled using a quantum variable, provided that one takes care in how one \tit{interprets} the result of the estimation. In particular, whether estimation can be consistently interpreted as revealing the physical quantity's true value becomes a live question that must be examined. 

On the conservative view, parameter estimation only makes sense when the estimand is modeled explicitly as a parameter. This does not by any means prevent us from doing estimation of quantities that are regarded as \tit{fundamentally} quantum, or which \tit{can} reasonably be modeled quantum mechanically in some circumstances --- rather, the conservative view simply asserts that one cannot perform estimation of the quantity until one is certain that the circumstances justify modeling the estimand as a parameter. In this respect, the conservative view can never fail to be consistent  --- the price it pays for consistency is a narrowing of scope.

In favour of the pragmatic view, we point to two facts. First, for the physical quantities we are concerned with here, the choice to model them as quantum variables or parameters is a matter of judgement on the part of the physicist, and so it appears essential to provide guidance as to when the predictions of the former model coincide with the predictions of the latter, thereby justifying the switch from a quantum variable to a parameter. The conservative approach essentially sweeps this problem under the rug, as it were, by simply presupposing all the justificatory work has already been done. For situations where the physicist may be uncertain about how to model the physical quantity, that approach is useless. 

Second, as mentioned in the Introduction, it is a simple mathematical fact that any model which treats a physical quantity as a parameter can be replaced by a model that makes identical predictions yet treats the quantity as a quantum variable. As we will see in Sec.~\ref{sec:q-control}, the abstract structure of an estimation protocol can be carried over directly to the new model, enabling one to provide a straightforward definition of ``estimation of a quantum variable" that is strictly more general than the classical model. Moreover, as we will see, the realist interpretation of estimation can also be carried over, allowing us to view the result of estimation as revealing a pre-existing value of a quantum variable, \tit{without} running afoul of the no-go theorems for hidden variables, such as Bell's theorem. Thus, where the conservative approach to estimation would put a blanket ban on estimation of quantum variables, the pragmatic view is more permissive, shifting the emphasis to the question of justificatory \tit{coherence}. It asks: under what assumptions is it \tit{consistent} to regard the process of estimation as informing us about the `true value' of a quantity, without thereby inadvertently positing a hidden variable? As we will see, in trying to answer this question, the pragmatic approach illuminates how quantum estimation is implicitly bound up with questions about the interpretation of quantum theory, establishing a hitherto unsuspected link between quantum metrology and quantum foundations.

\section{Circuit model of parameter estimation  \label{sec:framework}}

Quantum estimation protocols in the literature usually assume that the quantity to be estimated is a parameter. However, if we consider a physical quantity that is usually modeled as a quantum variable in the most fundamental theories in which it appears --- prominent examples being electric field magnitude, energy, angular momentum, \tit{etc} --- then our ability to model it as a parameter in an estimation protocol needs to be viewed as a special limit of the more general quantum model.

In the following sections we show how estimation of a quantity, initially viewed as a parameter, can always formally be embedded in a more general model in which the same quantity appears as a quantum variable.

To achieve this, we rely on the fact that any classical theory of a parameter $\theta \in \Theta$ can be formally represented as a strict sub-theory of a quantum system whose dimension equals the cardinality of $\Theta$, in precisely the same way that a classical bit can be viewed as the decoherent classical subtheory of a qubit in quantum information theory. The classical subtheory may be defined by restricting all states, transformations and measurements to be diagonal in the basis that designates the classical subspace. We shall make use of the converse statement: given a parameter $\theta \in \Theta$, it is always possible to \textit{embed} it into a quantum model by encoding it in the eigenstates of a suitable observable $\hat{\theta}$ whose eigenspectrum is isomorphic to $\Theta$. The more general model reproduces the predictions of the original, provided we restrict states, transformations and measurements accordingly.

The following two sections describe this generalized model; afterwards we will use it to investigate what goes wrong when the would-be estimand $\hat{\theta}$ is permitted to exhibit explicitly quantum behaviour.

\subsection{Classical controlled model \label{sec:c_control}}

For simplicity, we will focus on channel estimation with a single parameter using $N$ initially uncorrelated probes. While this is a highly restricted setting --- it excludes, for instance, many important protocols in which quantum entangled probes can provide a precision advantage over classical probes~\cite{escher_general_2011} --- it is entirely sufficient for the purposes of our argument. Indeed, since the only entanglement that is relevant to our argument is entanglement between the probes and the system being probed, we do not wish to confuse the reader by introducing additional entanglement between probe systems which plays no role in our argument.

In this restricted context, a classical or quantum estimation protocol may be divided into the following general stages:\\

\begin{enumerate}
    \item \tit{Preparation:} The system that hosts the physical quantity of interest, which we label system $A$, is prepared in a manner consistent with the assumption that $\theta$ has the same fixed (if unknown) value in every trial. The $N$ probe systems, labelled $B_1,\dots,B_N$, are independently prepared so as to be assigned identical and uncorrelated states;
    \item \tit{Imprinting:} The probes are submitted to a physical process that alters their state in a manner that depends on $\theta$, through physical interaction with the system $A$
    \item \tit{Measurement:} The probes are submitted to independent (possibly different) measurements $X_1,X_2,\dots,X_N$, resulting in measurement data $\vec{x} := \{x_1,\dots,x_N \}$. This data is used to estimate $\theta$.
\end{enumerate}

To give a simple example from classical estimation theory, suppose that one wishes to measure the melting temperature of sodium, represented as an unknown parameter $T$. In the preparation phase, a set of $N$ thermometers is brought to thermal equilibrium with some starting reference temperature (such as the ambient room temperature); during imprinting, the thermometers are submerged in the sodium at the moment of its melting; then during the measurement phase the indications on the thermometers are read and the temperature reading of each one is inferred, yielding the data $\vec{x}$. One can then use several strategies to produce an estimate $T_{\trm{est}}$: a simple strategy is to just average the thermometer readings, while a more complex strategy might involve a prior probability $P_0(T)$ over possible values of $T$ that is subsequently updated conditional on the data $\vec{x}$ using Bayes' rule. Choosing a good \tit{estimator} function $T_{\trm{est}} := f(\vec{x})$ is an important part of estimation, however in the present work we shall not be concerned with the particular choice of estimator --- any one will do.

In the context of \tit{quantum} estimation theory, the probe systems and the process that depends on the unknown parameter $\theta$ are modeled quantum mechanically; however it is important to note that $\theta$ itself is usually considered to remain a \tit{classical} parameter, that is, it is standard to assume that $\theta$ always has some determinate --- if unknown --- value, and never a quantum superposition of values. (Later, we will investigate what happens when this assumption is relaxed.)

Recall the example given previously of inferring some quantity $F$ of a magnetic field by estimating the precession angle $\eta$ that it imparts to a spin-$1/2$ atom used as a probe. Here, the parameter to be estimated is $F$, which refers to a property of the system $A$; the angle $\eta$ is a proxy for $F$, and is itself a parameter that describes a property of the probe system(s).

It is usually assumed that the quantum fluctuations of the magnetic field are too small to be important, hence values of $F$ cannot be in quantum superposition --- at most, it can be a classical probabilistic mixture of possible values. The field's action on the atom is modeled as a unitary operator $U_{F}$ that rotates the state vector around a fixed axis by an angle $\eta(F)$, that is,
\eqn{ \label{eqn:magnetic_classical}
U_{F} \ket{\psi_{\trm{in}}} = \ket{\psi_{\eta(F)}} \, .
}

\begin{figure}[h]
\centering
\includegraphics[width=8cm]{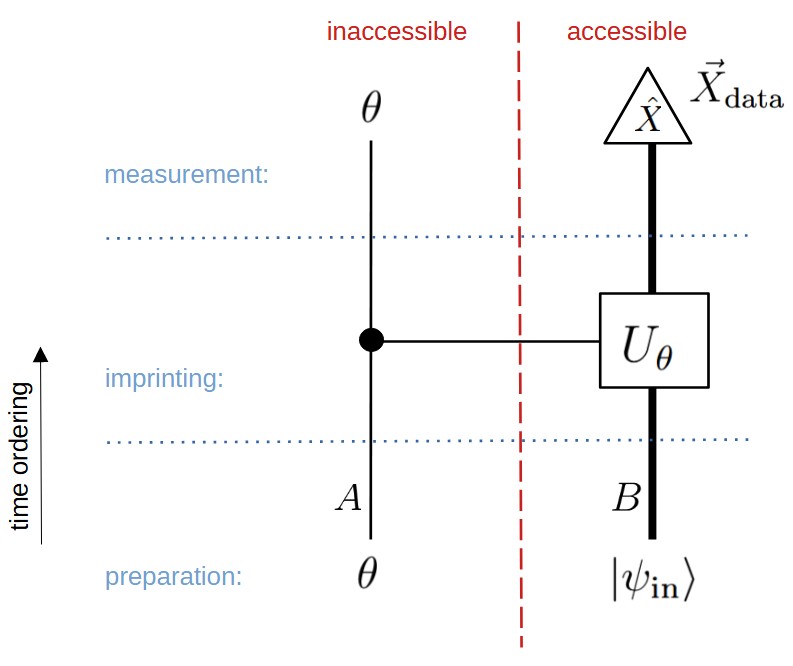}
\caption{ \label{fig:PE_circuit_simple} Quantum parameter estimation represented as a classically-controlled quantum circuit. The order of processes runs up the page; thin wires represent classical systems, while thick wires represent quantum systems. The unknown parameter $\theta$ is modeled as information stored in a classical register $A$, which serves as the classical control for a unitary process, $U_{\theta}$, however the system comprising the register is not directly accessible to the experimenter. The probe system $B$, initialized in the state $\ket{\psi_{\trm{in}}}_B$, is thereby subsequently `imprinted' with the value of $\theta$. By doing a measurement $X$ on the probe, one seeks to infer $\theta$.}
\end{figure}

A quantum parameter estimation protocol may therefore be represented by a simple quantum circuit as shown in Fig.\ref{fig:PE_circuit_simple}, for the case of a single probe. 

According to this model, for a given prior distribution $P_0(\theta)$ over the possible values of $\theta$, the quantum state of the $N$ probes after the imprinting phase has the general form:
\eqn{ \label{eqn:probe_out}
\rho^{\trm{out}}_B = \int \, P_0(\theta) \, \trm{d}\theta \, \rho^{\otimes N}_{\theta} \, , 
}
where $P_0(\theta) \, \trm{d}\theta$ is a measure on the set $\Theta$ and $\rho_{\theta}$ is the state of any given probe conditional on $\theta$:
\eqn{
\rho_{\theta} &=& \ketbra{\psi_{\theta}}{\psi_{\theta}} \nonumber \\ 
&:=& U_{\theta}\ketbra{\psi_{\trm{in}}}{\psi_{\trm{in}}} U^{\dagger}_{\theta} \, .
}
When considering $N$ probes, the Hilbert space of the probe systems decomposes as $\mathcal{H}_B := \mathcal{H}_{B_1} \otimes \dots \otimes \mathcal{H}_{B_N}$, the initial state of the probes is denoted $\rho^{\otimes N}_{\trm{in}}$, and the unitary acting on all $N$ probes is extended accordingly to $U_B(\theta) := U^{\otimes N}_{\theta}$.
 
Conceptually, the `classical register' that stores the value of $\theta$ models the physical degrees of freedom hypothesized to exist in the system $A$ that determine the value $\theta$.

By assumption, $\theta$ represents a physical quantity, and so it is meaningful to contemplate a hypothetical ideal measurement of $\theta$, represented in the circuit model as `reading the classical register $A$', that would in principle reveal its exact value. However, it is also assumed that performing such a measurement would be impractical, hence the need for estimation via the less direct route of measuring the probe systems. Thus, we have labeled $A$ `inaccessible' and the probes $B$ as `accessible' in the figure.

For the case of melting sodium, for instance, the individual molecules of sodium may be assumed to possess a particular distribution of velocities at the moment of melting, whose mean value in principle determines the unknown temperature $T$ prior to and independently of any attempt to measure it.

In the example of probing a magnetic field with an atom, the magnetic field itself --- modeled classically --- is assumed to possess a definite value of the component $F$, which could in principle be discovered by a hypothetical sufficiently accurate measurement of the field at all relevant points, and which, if known, would perfectly determine the precession angle $\theta$ imparted to the quantum probe.

\subsection{Quantum-controlled model \label{sec:q-control}}

\begin{figure}[h]
\centering
\includegraphics[width=8cm]{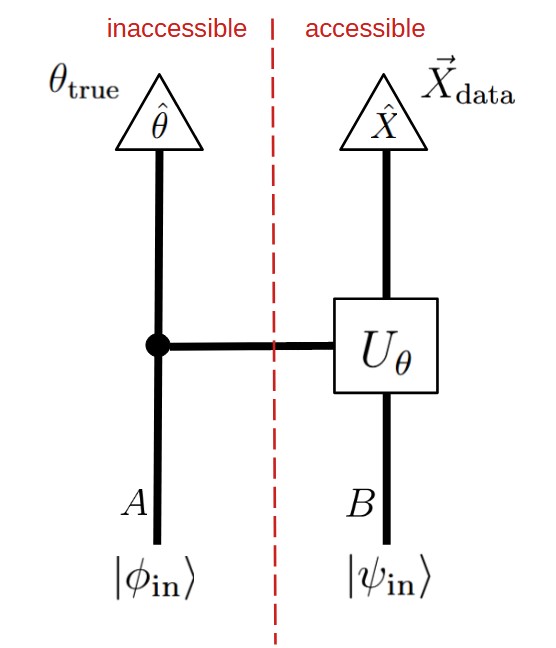}
\caption{ \label{fig:PE_circuit_qm} Quantum parameter estimation represented as a quantum-controlled circuit. The assumed inaccessible physical quantity of interest is now modeled quantum mechanically (thick wire); it becomes quantum entangled to the probe systems via the ``quantum controlled'' process (control with thick wire). The output is an entangled state with no clearly defined value for the quantity. A hypothetical final measurement of the observable $\hat{\theta}$, if performed, would produce a ``true'' parameter value $\theta$ consistent with the measurement data $\vec{X}_{\trm{data}}$ from the probe systems.}
\end{figure}

We now show that the circuit model for estimation of a parameter $\theta$ described in the previous section can be fully embedded in a quantum model that reproduces the predictions of the former, yet represents the quantity of interest by an observable $\hat{\theta}$. A schematic of the more general model is shown in Fig.~\ref{fig:PE_circuit_qm}. 

The physical degrees of freedom $A$ are now modeled as a quantum system\footnote{This method of embedding classical information into the quantum formalism by means of a preferred basis is standard practice in quantum information (the bit/qubit correspondence being the paradigmatic case), and has been employed in the quantum metrology literature before --- see for instance~\cite{zhang_molmer_2020}.} whose Hilbert space $\mathcal{H}_A$ is spanned by an orthonormal basis labeled by the possible eigenstates of an observable $\hat{\theta}$, whose eigenvalues are identical to the values of the parameter $\theta$. Formally, if the set $\Theta$ is discrete, the basis is defined by the (finite or countably infinite) set of eigenstates $\{ \ket{\theta} : \theta \in \Theta\}$, where $|\braket{\theta_i}{\theta_j}|^2 = \delta_{ij}$. In this case, a hypothetical measurement of $ \hat{\theta}$ would be an \tit{ideal} measurement, in the sense that it can perfectly distinguish any two distinct eigenstates of the basis without error. 

Note that the dimension of $\mathcal{H}_A$ equals the cardinality of the set $\Theta$; in particular if $\theta$ is a continuous parameter as usually assumed, then $\trm{dim}(\mathcal{H}_A)$ is infinite.
In order to consider such parameter spaces that are continuous, however, we encounter a minor technical problem: the basis associated with a continuous spectrum $\theta \in \Theta$ satisfies $|\braket{\theta}{\theta'}|^2 = \delta(\theta-\theta')$ where $\delta$ is the Dirac functional; consequently the eigenstates are not normalizable. This is expected, since no realistic measurement could pinpoint $\theta$ to an arbitrary number of decimal places; however it means that we cannot use the eigenstates of $\hat{\theta}$ to encode information about $\theta$ into the system $A$, because they cannot be normalized and so cannot represent valid states of a physical system. 

The solution is to define a more realistic hypothetical measurement on $\mathcal{H}_A$ that is defined in terms of projections onto `smeared-out' states $\ket{f_{\theta'}} := \int \, \trm{d}\theta \, f_{\theta'}(\theta)  \ket{\theta}$. The smearing function $f_{\theta'}(\theta)$ is normalized, $\int |f_{\theta'}(\theta)|^2 \trm{d}\theta = 1$, and sharply peaked at some value $\theta'$; its particular shape depends on a detailed model of the measuring instrument. In particular, the variance of $f_{\theta'}(\theta)$ quantifies the minimum spacing needed between distinct values of $\theta$ in order for them to be distinguished by the measurement with high confidence. This motivates us to replace the (unphysical) notion of an ideal measurement with the notion of a \tit{sufficiently accurate} measurement:\\

\tbf{Definition:} Let $\epsilon >0$ be a real number that quantifies an in-principle limit to the precision with which $\theta$ can be determined, either due to physical principles or vagueness in the definition of $\theta$ itself; and consider the Hilbert space defined mathematically by the orthogonal basis of (unphysical) states $\{ \ket{\theta} \}$ with eigenvalues drawn from a continuous set $\Theta$. A \tit{sufficiently accurate} quantum measurement of $\theta$ comprises (i) a partition $\mathcal{J}$ of $\Theta$ into non-overlapping `bins' of finite size $\theta_j \pm \epsilon$, with center points $\theta_j$ labeled by a discrete index $j$, and (ii) an associated set of projection operators $\{ \ketbra{f_{\theta_j}}{f_{\theta_j}} \}_{\forall j}$ which are normalized and mutually orthogonal, i.e.\ $ |\braket{f_{\theta_j}}{f_{\theta_k}}|^2 = \delta_{jk}$. Given any value $\theta' \in \Theta$, we shall let $[\theta'] \in \mathcal{J}$ denote the range of values of the bin which contains $\theta'$. \\

We stress that while our notion of sufficiently accurate measurement is somewhat more realistic than the ideal measurement, it is still very much a \tit{hypothetical} measurement that represents the in-principle limit of precision, rather than any practical limit. Indeed, if we are permissive regarding what physical laws may allow, then $\epsilon$ can be extremely small. Ultimately, depending on what it represents, it may not even be limited by physical principles at all, but only by semantics: the bin size cannot meaningfully be smaller than the `definitional uncertainty' of the quantity of interest, defined in the VIM 3rd edition \cite{VIM_2012}, Section 2.27, as: ``component of measurement uncertainty resulting from the finite amount of detail in the definition of a measurand.'' This is entirely appropriate here, since the role played by the sufficiently accurate measurement in our argument is essentially definitional: we want to have the option of being able to \tit{define} the `true value' of a quantity as the value that would be obtained by a sufficiently accurate measurement of it (cf.\ the \tit{Independence} assumption). In this case, the limit of precision cannot be attributed to imperfections of the instrument; it can only come from vagueness in the very definition of the quantity itself. 

Using this definition, we can embed a continuous parameter $\theta \in \Theta$ into a \tit{discrete} Hilbert space by the following procedure. We first construct the `provisional' infinite-dimensional Hilbert space $\tilde{\mathcal{H}}$ from the continuum of unphysical eigenstates $\ket{\theta}$ as described above, then introduce a sufficiently accurate measurement $\{ \ketbra{f_{\theta_j}}{f_{\theta_j}} \}_{j \in \mathcal{J}}$. The states $\{\ket{f_{\theta_j}}\}_{j \in \mathcal{J}}$ define an orthonormal basis for a \tit{discrete} subspace $\mathcal{H}_{\mathcal{J}} \subset \tilde{\mathcal{H}}$. On this subspace the observable defined as $\hat{\theta}_{\mathcal{J}} := \sum_j \theta_j \ketbra{f_{\theta_j}}{f_{\theta_j}}$ is therefore physically allowed and it can perfectly distinguish any two states that have support in different bins. We then propose to embed $\theta$ into the subspace $\mathcal{H}_{\mathcal{J}}$ by encoding each value $\theta$ as the eigenstate corresponding to the bin which contains it:
\eqn{
\{ \theta \mapsto \ket{f_{\theta_j}} : [\theta] = [\theta_j] \}_{\theta \in \Theta} \, .
}
Note that this is not a faithful embedding, since it is many-to-one: all values of $\theta$ that lie within the same bin get mapped to the same quantum state. Thus, we may replace the continuous set $\Theta$ with the discrete set $\Theta / \mathcal{J}$ comprising just the center points $\theta_j$ of the bins. We stress that no generality is actually lost by this procedure provided we choose the sufficiently accurate measurement to have a bin size comparable to the definitional uncertainty\footnote{We acknowledge that this is something of an idealization of how language works, since the `size' of definitional uncertainty is also not something that can be precisely defined. As long as there is some plausible definition of $\theta$ that is blind to differences of the order of $\epsilon$, we consider this an adequate choice for the purposes of defining the sufficiently accurate measurement.}. In that case, no meaningful information is lost by this procedure, since values within the same bin cannot be meaningfully distinguished even in principle. The price to be paid is that such an embedding is not unique, since it depends on the choice of sufficiently accurate measurement and the associated binning; however this ambiguity is not any more consequential than the ambiguity inherent in the definition of $\theta$ itself. 

Since discrete and continuous parameters alike are ultimately embedded into a discrete quantum model, we shall use the same notation for both cases. Thus, we will drop the subscript $\mathcal{J}$, letting it be understood implicitly in the case of a continuous parameter that $\mathcal{H}_A$ refers to the discrete subspace $\mathcal{H}_{\mathcal{J}}$; $\hat{\theta}$ refers to $\hat{\theta}_{\mathcal{J}}$; $\theta \in \Theta$ refers to $\theta_j \in \Theta / \mathcal{J}$; and $\ket{\theta}$ refers to $\ket{f_{\theta_j}}$. Finally, when discussing a discrete embedding of a continuous parameter, a measurement result of `$\theta_j$' should not be interpreted as identifying a unique value, but rather indicates a range of values consistent up to definitional uncertainty (cf.\ the definition of \tit{Uniqueness} later on).

Having now defined the embedding of the parameter into a quantum model, the transformation $U_{\theta}$ on system $B$ is now accordingly upgraded by extending it to the joint Hilbert space $\mathcal{H}_A \otimes \mathcal{H}_B$. Given a preferred basis (here, the eigenstates of $\hat{\theta}$) this extension is uniquely defined by 
\eqn{
U_{AB}(\ket{\theta} \otimes \ket{\psi}) := \ket{\theta} \otimes (U_B(\theta)\ket{\psi}) \, ,
}
recalling that $U_B(\theta) := U^{\otimes N}_{\theta}$. This is the standard form of a `quantum controlled unitary', whose functioning may be interpreted as follows (see Fig.~\ref{fig:PE_circuit_qm}). Data is encoded in the `control basis' of a \tit{quantum register} --- here, the basis of $\hat{\theta}$ eigenstates of the system $A$ --- which then serves as a \tit{control} that determines which unitary $U_{\theta}$ is applied to system(s) $B$. Provided the initial state of the control system $A$ is diagonal in the $\hat{\theta}$-basis, the circuit functions exactly in the same way as the classically controlled circuit depicted in Fig.~\ref{fig:PE_circuit_simple}, thereby recovering the predictions of the model in which $\theta$ is a parameter. The following two propositions formalize this statement. The first asserts that, under the `diagonality' constraint, the final state of the probe systems (when considered alone) is unaffected. \\

\tbf{Proposition 1:} Suppose the initial state of $A$ is diagonal in the control basis, that is,
\eqn{
\sigma_{\trm{in}} = \zum{\theta \in \Theta}{} \, P(\theta) \, \ketbra{\theta}{\theta} \, ,
}
with $P_0(\theta)$ a probability measure over $\Theta$. Then, according to the quantum controlled model, the output state of the $N$ probes after imprinting is given by:
\eqn{ \label{eqn:rho_out_discrete}
\rho^{\trm{out}}_B = \zum{\theta \in \Theta}{}\, P_0(\theta) \,  \, \rho^{\otimes N}_{\theta} \, , 
}
which is readily seen to be equivalent to the discrete form of the output from a classical controlled circuit~\eqref{eqn:probe_out}. (In case the parameter is continuous, equivalence is obtained in the limit as the bin size goes to zero, $\epsilon \rightarrow 0$, allowing us to replace $\sum_{\theta} \mapsto \int \trm{d}\theta$ in the formula.)\\

\tit{Proof:} Given the input state $\sigma_{\trm{in}} \otimes \rho^{\otimes N}_{\trm{in}}$, the joint state of systems $A,B$ emerging after the imprinting stage is:
\eqn{
\rho^{\trm{out}}_{AB} &=& U_{AB}(\sigma_{\trm{in}} \otimes \rho^{\otimes N}_{\trm{in}})U^{\dagger}_{AB} \nonumber \\
&=&  \zum{\theta \in \Theta}{} \, P_0(\theta) \,   \,  U_{AB}\left( \ketbra{\theta}{\theta} \otimes \rho^{\otimes N}_{\trm{in}} \right) U^{\dagger}_{AB}  \nonumber \\
&=&  \zum{\theta \in \Theta}{} \, P_0(\theta) \,   \,  \ketbra{\theta}{\theta} \otimes \left( U_{\theta} \, \rho_{\trm{in}} \,  U^{\dagger}_{\theta} \right)^{\otimes N} \nonumber \\
&=&  \zum{\theta \in \Theta}{} \, P_0(\theta) \,   \,  \ketbra{\theta}{\theta} \otimes  \rho^{\otimes N}_{\theta} \, .
}
The output state of the probes is then obtained by taking the partial trace, $\rho^{\trm{out}}_B = \trc{A}{\rho_{AB}}$, which by inspection gives~\eqref{eqn:rho_out_discrete}. $\Box$

The preceding result only concerns the final state of the probes; however it is equally important to consider the correlations that are established between the probe systems and quantity of interest as represented by the states of system $A$. Since the latter are now also represented quantum mechanically, it is natural to wonder whether the unitary interaction between $A$ and $B$ might cause them to become quantum entangled to one another. The following Proposition establishes that the answer is negative, if the initial state of $A$ is diagonal in the control basis:\\

\tbf{Proposition 2:} If $\rho^{\trm{out}}_{AB}$ is entangled, then $\sigma_{\trm{in}}$ must be initially assigned a quantum state that is not diagonal in the control basis.\\

\tit{Proof:} We proceed by contradiction: suppose $\sigma_{\trm{in}}$ were diagonal in the control basis: 
\eqn{
\sigma_{\trm{in}} = \zum{\theta \in \Theta}{} \, p_{\theta} \ketbra{\theta}{\theta} \, ,
}
for some probabilities $p_{\theta}$. Applying $U_{AB}$ to the initial state of $\sigma_{\trm{in}} \otimes \rho_{\trm{in}}$ we obtain the joint state after interaction:
\eqn{
\sigma^{\trm{out}}_{AB} = \zum{\theta \in \Theta}{} \, p_{\theta} \ketbra{\theta}{\theta} \otimes \rho_{\theta}\, .
}
This state meets the definition of a \tit{separable} state between the subsystems $A$ and $B$, meaning (by definition) there is no entanglement between them. On the other hand, the Entanglement assumption asserts that $\rho^{\trm{out}}_{AB}$ \tit{is} entangled between $A$ and $B$. We conclude that the initial state $\sigma_{\trm{in}}$ must not be diagonal in the control basis. $\Box$ 

Based on the above two Propositions, the decision to `lift' the classical controlled model of estimation to a quantum controlled model by embedding the parameter in a Hilbert space does not --- under the diagonality constraint --- imply any alteration to the predictions of the model for the results of measurements on the probe systems, nor any alteration to how those results are correlated with the value assigned to the quantity of interest. Under this constraint, the two models are identical for all practical purposes, and the only difference is a purely formal matter of how $\theta$ is represented within the models. 

Consequently, in this restricted case the interpretation of the classical model may be carried over to the quantum model. Specifically, the value one would obtain from a sufficiently accurate measurement of $A$ may safely be interpreted as equivalent to the \tit{true value} of the parameter (up to definitional uncertainty), and the model predicts that --- for an unbiased estimator --- the estimated value converges to this true value in the large-$N$ limit.

To gain some intuition about the quantum controlled model, let us apply it to the example of estimating magnetic field direction. Although the magnetic field is considered classical, we may still model it using the quantum mechanical formalism by introducing a quantum observable $\hat{F}$ that represents a sufficiently accurate measurement of the relevant component of the field, $F$. Given a classical solution to the field equations for this component\footnote{For clarity of exposition we have suppressed the field's dependence on space-time co-ordinates $(x,t)$; for the purposes of our example we may regard the system $A$ as comprising the field at some designated point in space and at some moment in time just before the imprinting stage.}, we may simply define the quantum state of $A$ to be $\ket{F}$ such that $\hat{F} \ket{F'} = F' \ket{F'}$, that is, an eigenstate whose eigenvalue picks out the unique value corresponding to the classical solution $F=F'$. Thus, in this example we may take the eigenstates of the observable $\hat{F}$ as defining an orthonormal basis for the (discrete) Hilbert space $\mathcal{H}_A$. Each probe system is identically prepared in a state with fixed angular momentum, denoted $\ket{\uparrow_{0}}$; the initial $N$-probe state is $\rho^{\otimes N}_{\trm{in}} = ( \ketbra{\! \uparrow_{0  }}{\uparrow_{0  } \! })^{\otimes N}$. The unitary interaction with the system $A$ causes each probe to rotate by a fixed angle $\phi(F)$ that depends on the field, thus $U_{AB}\ket{F}\ket{\!\! \uparrow_{0  }} =
\ket{F}\ket{\!\! \uparrow_{\phi(F)  }}$. For a fixed value of $F$, the probes emerge from the interaction with state $\rho^{\otimes N}_{\phi(F)} := (\ketbra{\!\! \uparrow_{\phi(F)  }}{ \uparrow_{\phi(F)  }\!\!})^{\otimes N}$. More generally, assuming the initial state of $A$ is diagonal in the $\hat{F}$ basis, with prior distribution $P_0(F)$ over the possible values of the field, the output state of the probes is 
\eqn{
\zum{F}{} \, P_0(F) \, \rho^{\otimes N}_{\phi(F)} \, ,
}
which, following the classical model, may be interpreted as expressing the belief that the probe states are all aligned along some definite angle $\phi(F)$ corresponding to some determinate field value, and $P_0(F)$ expresses our uncertainty about what this true value is. Given that a hypothetical sufficiently accurate measurement of $\hat{F}$ would yield the result $F'$ (accurate to within definitional uncertainty), we should then expect that in the limit of large $N$, the estimation procedure should converge to a value that lies within the same bin, which can be interpreted as revealing the pre-existing true value of the field magnitude.

The question raised by this more general model is what goes wrong with the above interpretation when we stray beyond the confines of the classical subtheory by allowing the initial state of system $A$ to be a superposition of distinct eigenstates, i.e., non-diagonal in the control basis. The following section examines the assumptions that underwrite the classical interpretation, and shows how allowing quantum superpositions of the estimand places them in tension with certain commitments about the interpretation of quantum theory itself.

\section{A No-Go Theorem for True Values in Quantum Estimation \label{sec:problem_QPE}}

In what follows, we will isolate five assumptions that become salient when one attempts to perform estimation of a quantum variable. In brief, the first three assumptions --- \tit{Uniqueness, Pre-existing Value, Independence} --- apply to both classical and quantum estimation. They express what we take to be the minimal realist content of estimation: that there is a value to be estimated, that the value is not merely created by the measurement, and that the value is linked to what a sufficiently accurate (or ideal) measurement would reveal.

The final two assumptions --- \tit{Entanglement, Completeness} --- are distinctively quantum. They express the thought that, first, the quantity of interest may become entangled with the probes, and second, given a system assigned a pure quantum state, there should not be any additional facts (i.e.\ `hidden variables') that could allow one to predict the outcomes of measurements with greater precision than when using the quantum state alone.

Our claim is that these two sets of assumptions are jointly incompatible: either one rejects one of the quantum assumptions, or else a realist account of estimation is impossible.

We begin by briefly reviewing important terms and concepts established in previous sections. As usual, $\Theta$ denotes the set of \textit{ab initio} possible values of $\theta$; we shall also find it useful to use primes, $\theta', \theta''$, \textit{etc.}, to denote distinct values in $\Theta$ that are perfectly distinguishable by a sufficiently accurate measurement, hence separated by at least the minimum bin size. We also remind the reader that $[\theta']$ indicates the range of values within the bin that contains $\theta'$.

We consider an estimation protocol involving a fixed number of probes $N$; after all probes are measured and an estimate of $\theta$ is produced, the protocol ends and no further probes are used. Each probe's preparation, interaction with the system $A$, and measurement are independent of all the other probes. Consequently, it does not matter for our analysis whether they are measured in sequence or all together. We will find it convenient to use the latter convention, treating all the probes as a single joint system, denoted $B$.   

Consider now the following assumptions:\\

\noindent \tbf{Uniqueness:} At the completion of the estimation protocol, given any finite set of distinguishable hypotheses about the value of $\theta$ of the form $\theta \in [\theta']$, for values in a finite set $
\{ \theta', \theta'', \dots
\} \subseteq \Theta$, all of which are consistent with the measurement data, \tit{exactly one} of them is in fact true and the rest are false\footnote{It is expected that in the limit of large $N$ the set of distinct hypotheses consistent with measurement data should shrink to just one element; but since we aim to cover cases where $N$ might be small, or even 1, we allow that there may be multiple hypotheses consistent with the data.};
\\

\noindent \tbf{Pre-existing Value:} A given hypothesis $\theta \in [\theta']$ is true when the measurements are completed if and only if it was already true \tit{before} the measurements were carried out;\\

\noindent \tbf{Independence:} The hypothesis $\theta \in [\theta']$ is true at a given time if and only if a hypothetical \tit{sufficiently accurate measurement} of $\theta$, if it were performed at that time, \tit{would} produce the result $\theta'$.\\

The next two assumptions refer explicitly to quantum theory, and as such are specific to quantum parameter estimation:\\

\noindent \tbf{Entanglement:} The process of interaction between the probes $B$ and the system $A$ leads to their joint state $\rho^{\trm{out}}_{AB}$ being an \tit{entangled state};\\ 

\noindent \tbf{Completeness:} If a hypothetical sufficiently accurate measurement of $\theta$ on system $A$ would produce the result $\theta'$ with certainty at a given time, then the quantum state of $A$ must be consistent with this fact. Formally, the state of $A$ must be expressible as a probabilistic mixture of eigenstates of $\hat{\theta}$ that includes the eigenstate $\ket{\theta'}$. \\ 

\tit{Remark:} 
Proposition 2 tells us that by accepting the Entanglement assumption, we are thereby accepting that the state of $A$ is initially \textit{not} diagonal. This amounts to asserting that any decoherence process which is operative via the coupling of system $A$ to environmental degrees of freedom is only partial: it suppresses the off-diagonal terms, but does not fully eliminate them. We note that this is a generic feature of realistic models of decoherence~\cite{Zurek2003}. Thus, for the quantities that we are considering here --- those which may reasonably be modeled as quantum variables --- Entanglement is expected to hold. 
To dispel any skepticism that there can be any realistic experimental situation that meets these conditions, consider the case of using an atomic probe (system $B$) to estimate the magnitude of some component of a magnetic field, which we assume is generated by a current circulating in a conductor (system $A$). The estimand can be modeled as a parameter only when quantum fluctuations of the field can be ignored. But can they always be ignored? An experiment recently proposed in Ref.~\cite{raman_nair_2025} suggests using superconducting flux qubits to create quantum superpositions of current flowing in opposite directions, leading to a quantum superposition of magnetic fields. To the extent that one can control and measure the current in the flux qubits, one can in principle measure the system $A$, potentially detecting entanglement between the magnetic field and the probe systems $B$ that interact with it; in this case it would be quite inappropriate to treat the field magnitude as a parameter.

Since decoherence cannot be relied upon to rule out Entanglement in every situation, rejecting this assumption requires something stronger; we will discuss some possibilities in Sec.~\ref{sec:de_finetti} and Sec.~\ref{sec:implications}.

\tit{Remark:} Completeness is important because it provides a link to the concept of hidden variables. To see how, consider what it would mean to reject Completeness. Specifically, suppose a hypothetical sufficiently accurate measurement of $\theta$ on system $A$ would produce the result $\theta'$ with certainty at a given time, and suppose (contra Completeness) $A$ is assigned a state non-diagonal in the control basis. It suffices to consider the special case where the state of $A$ is pure, $\sigma_{\trm{in}} := \ketbra{\phi}{\phi}$. 

The state $\ket{\phi}$ must have nontrivial overlap with $\ket{\theta'}$ or else it is impossible that a sufficiently accurate measurement would yield $\theta'$; on the other hand if it is non-diagonal then it must be a pure superposition involving at least one other basis element, let us say $\ket{\theta''}$. Thus the state has the general form:
\eqn{
\ket{\phi} = \alpha' \ket{\theta'} + \alpha''  \ket{\theta''} + \dots  \,,
}
where $\alpha',\alpha''$ are some nonzero complex amplitudes and `$\dots$' allows for additional unspecified terms. This is a pure superposition that predicts the outcome $\theta''$ with probability $|\alpha''|^2$ when measuring $\hat{\theta}$, so it implies that there is some fact about $\theta$, namely that $\theta \in [\theta']$, which is not represented in the quantum state of $A$. That is to say, the state $\ket{\phi}$ is \tit{incomplete}; completing it would require supplementing it with an additional variable that would let us narrow the possibilities down to $\theta \in [\theta']$. Since $\ket{\phi}$ is pure, this supplemental variable would moreover constitute a ``hidden variable'' by the definition used in Bell's theorem, that is, a variable that determines the outcome of a possible measurement, yet whose value is not determinate even for a pure quantum state. Thus, if one wishes to reject hidden variable accounts of quantum theory of the kind targeted by Bell's theorem, then one should accept Completeness. \\

\tit{Theorem:} The set of assumptions listed above are mutually inconsistent.\\

\tit{Proof:} Uniqueness requires that exactly one hypothesis be true at the completion of the estimation; call it $\theta'$. Then Pre-Existing Value implies that $\theta'$ must also have been the unique true hypothesis at the initial time, i.e.\ before interaction with the probes. It follows from Independence that a hypothetical sufficiently accurate measurement of $A$ at the initial time would have resulted in a value $\theta'$, and Completeness then implies that the state of $A$ is diagonal in the $\hat{\theta}$-basis. On the other hand, Proposition 2 proved that if Entanglement holds, the initial state of $A$ \tit{cannot} be diagonal in this basis --- a contradiction. $\Box$ \\

\section{Discussion \label{sec:discussion}}

\subsection{Connection to Bell's Theorem \label{sec:Bell}}
In discussing the assumption of Completeness, we noted its connection to the concept of hidden variables. Thus, our main Theorem proved in the preceding section can be read as a species of no-go Theorem for hidden variables. Namely, our theorem implies that the only way to maintain a realist view of estimation in the presence of entanglement between the probes and the estimand is to accept hidden variables (i.e.\ because one must then reject Completeness). In this section, we make this connection explicit by discussing how our theorem relates to the well-known hidden variable theorems in the literature, focusing on the EPR argument and Bell's theorem.
 
We begin with a brief review. The 1935 `EPR argument' of Einstein, Podolsky and Rosen was arguably the first important result concerning `hidden variables'~\cite{einstein1935can}. The authors leveraged a well-known feature of quantum entanglement between two widely separated particles: by measuring the position (respectively momentum) of one particle and conditioning on the outcome, the quantum state of the remaining distant particle should be updated to a position (respectively momentum) eigenstate, thus implying certitude about the value of the respective property. Assuming the inherent properties of the unmeasured particle (whatever they may be) would be undisturbed by such information gathering on the first particle --- quite reasonable, given that the distance between them could be made arbitrarily large --- it is then natural to posit that the unmeasured particle \tit{already} possesses sharp (i.e.\ determinate) values of \tit{both} properties (this postulate is known as the \tit{EPR reality criterion}). Since no quantum state permits position and momentum to be simultaneously sharply defined, the authors concluded that the quantum state must be incomplete, as a more complete description should posit `elements of reality' sufficient to uniquely determine both the particle's position and momentum in the situation described. (In hindsight, we see that such elements of reality would constitute hidden variables, though they were not called such at the time).

The argument sparked a debate, with many physicists --- notably Bohr --- disagreeing with the conclusion and choosing to reject one or another of its presuppositions. However, it was more than three decades before Bell's theorem proved definitively that hidden variables, if they do exist, must transgress at least one of several assumptions deeply entrenched in physics. 

Bell's 1976 theorem considered a scenario similar to that of EPR, namely, two particles widely separated in space upon which independent measurements could be performed. Bell proposed to explain the resulting measurement statistics using a general model of hidden variables, upon which only very weak requirements were imposed. Loosely speaking\footnote{To be precise, Bell proved two distinct theorems, one in 1976 and the other in 1964; moreover there is some flexibility in terms of how one decomposes his assumptions into more basic ones. Our presentation is based on the formulation of the 1976 theorem as given in Theorem 4 of Ref.~\cite{wiseman_causarum_2017}; our two loosely-described assumptions refer to \tit{Local Causality} and \tit{Local Agency} as defined more carefully in that article.}, he assumed that information about physical quantities could propagate no faster than the speed of light, and that if any factors existed that might plausibly affect the experimenter's choice of which measurements to perform, these factors would be wholly independent of the hidden variables that served to determine the particle's behaviour under measurements. His Theorem proved that all hidden variable models satisfying these requirements should satisfy a \tit{Bell inequality}: a formal restriction on the degree of statistical correlations allowed between the independent measurements. Finally, he exhibited a counterexample in which two quantum \tit{entangled} systems could in principle violate his inequality. In this manner, Bell went beyond EPR's essentially qualitative analysis of two-particle entanglement, providing a \tit{quantifiable} benchmark for the existence of a reasonable explanation for the correlations in terms of hidden variables. 

In the many decades after it was first proposed, Bell's inequality has been shown to be violated in a series of experiments of increasing stringency, involving pairs of entangled quantum systems separated by many kilometers, and culminating in three famous ``loophole-free'' experiments completed in 2015\footnote{For an accessible review with references to the literature, see Ref.~\cite{bell_tests}. For a more technical review aimed at philosophers, see Ref.~\cite{sep-bell-theorem}.}. The 2022 Nobel prize was awarded to Zeilinger, Clauser and Aspect, in no small part for their contributions to this experimental achievement.

Significantly for our discussion, in deriving his inequality Bell also made use of an \tit{implicit} assumption that later came to be known as \tit{counterfactual definiteness} (CFD), which states that conditional propositions of the form: ``If measurement $M$ had been performed, result $R$ would have been obtained'' always have a definite truth value, even when the measurement in question is \tit{not} carried out\footnote{Bell did not \tit{explicitly} state CFD. In his 1964 theorem, he assumes `locality' plus `determinism'; the latter turns out to be equivalent to assuming CFD~\cite{zukowski_2014}. In his 1976 theorem, he assumes `local causality', which alone does not imply CFD; however if one adds to this the QM prediction that entangled states exhibit perfect correlations, then CFD is implied. In this sense, Bell's theorem(s) presupposes CFD~\cite{kooderi_suresh_thesis}.}. Note that the `truth values' of CFD necessarily constitute hidden variables, since no pure quantum state implies determinate truth values for all observables. In fact, it is strictly stronger than EPR's reality criterion\footnote{As with Bell's Theorem, CFD is not explicitly stated in EPR's original formulation, but was later recognized as playing a key role~\cite{zukowski_2014}.}, because it implies hidden variables for any quantum state, not just eigenstates of position or momentum.

Turning now to a comparison with the present work, we note that both the EPR reality criterion and CFD bear a distinct resemblance to our Independence assumption, which also makes reference to the results of hypothetical measurements that need not actually be performed. In order to make the comparison cleanly, let us define \tit{counterfactual definiteness} in our terminology as the following: \\

\tbf{Definition:} \tit{Counterfactual definiteness (CFD)} states that if a hypothetical sufficiently accurate measurement of $\theta$ would result in the outcome $\theta'$ with certainty if it were performed, then there is a unique true hypothesis $\theta \in [\theta']$ at that time, \tit{even if some other measurement is actually performed}. \\

Thus, within the quantum-controlled model, we say $\theta$ is counterfactually definite at a given time if (i) there is a unique true hypothesis $\theta \in [\theta']$, and (ii) a hypothetical measurement of the observable $\hat{\theta}$, if it \tit{were} performed, would result in the outcome $\theta'$ with certainty. 

CFD is only interesting, of course, when the quantity that could be measured is actually \tit{not} measured, for only then does it lead to non-trivial consequences. Both the EPR argument and Bell's theorem apply it to the case of two alternative measurements on each particle, where the measurements are of distinct, `incompatible' observables (i.e.\ there is no quantum state that is an eigenstate of both observables), position and momentum being the paradigmatic example.

Here we see a clear contrast with our argument. Quantum parameter estimation sometimes makes use of measurements in incompatible observables (multiparameter estimation being a case in point), however, in the simplest protocols like the case we are considering, namely single-parameter estimation using uncorrelated probes, it is possible to perform estimation by always performing the same measurement on every probe. Indeed, our Theorem assumes throughout that the only measurements that could be performed are $\hat{\theta}$ on system $A$ and a generic, fixed and identical measurement denoted $\hat{X}$ on each probe system. Thus, to derive our contradiction we never need to contemplate distinct incompatible measurements on any sytem.

Having said that, our argument does rely on a \tit{variant} of counterfactual definiteness, as we now explain. We note that instead of contemplating a counterfactual measurement that is incompatible to the one actually performed, we could instead contemplate a counterfactual measurement of the \tit{same} observable but at an \tit{earlier time} than the time at which the measurements are actually performed.

This motivates the following definition: suppose that estimation of $\theta$ is completed based on measurements on the probe systems performed jointly at time $t_2$. Now consider what would have happened if, prior to these measurements, we had already done a sufficiently accurate measurement of $\theta$ at some earlier time $t_1 < t_2$, just prior to the interaction phase.\\

\tbf{Definition:} \tit{counterfactual definiteness in the past:} if an `early' measurement of $\hat{\theta}$ at time $t_1$ \tit{would have} resulted in the outcome $\theta'$, then there is a unique true hypothesis $\theta \in [\theta']$ at time $t_1 < t_2$ --- even when the early measurement was not actually performed. \\

Based on this definition, we can establish the link to our assumptions via the following result:

\tit{Claim:} The conjunction of Uniqueness \& Pre-existing Value \& Independence together imply that the quantity represented by $\theta$ is \tit{counterfactually definite in the past}, that is, prior to the estimation.

\tit{Proof:} Consider the system $A$ that is assumed to carry the quantity represented by $\theta$. As we saw in the proof of the main Theorem, the conjunction of Uniqueness \& Pre-existing Value imply that there is a unique true hypothesis $\theta \in [\theta']$ associated to system $A$ just prior to interaction with the probes. Adding Independence then establishes that a hypothetical sufficiently accurate measurement of $\theta$, if it \tit{had been} performed at the earlier time, would have resulted in the outcome $\theta'$ with certainty. (By assumption, of course, no such measurement is actually performed). $\Box$ \\

Finally, it is straightforward to see that the conjunction of the two assumptions motivated by quantum theory, namely Entanglement \& Completeness, are incompatible with counterfactual definiteness in the past. To see why, note that if the latter holds at a time just before the imprinting phase, then Completeness requires the state of $A$ to already be diagonal in the basis of $\hat{\theta}$; but as already noted above, this conflicts with Entanglement, which requires the initial state of $A$ to be non-diagonal in this basis.

The preceding discussion shows that the fundamental tension in our argument, as in Bell's theorem, can be construed as arising from a basic tension between counterfactual definiteness on one hand and quantum theory on the other. 

\subsection{Entanglement and the de Finetti constraint \label{sec:de_finetti}}

Among the set of inconsistent assumptions, Entanglement is the one that is likely to appear most suspect to practitioners of quantum parameter estimation, since they are not accustomed to modeling the estimand as a quantum variable; therefore the suggestion that the estimand could possibly become \textit{entangled} to the probe systems may strike them as strange, if not absurd. While it seems to us indisputable that experimental situations can be contrived in which the standard application of quantum theory would imply Entanglement (cf.\ the remark concerning Entanglement in Sec.~\ref{sec:problem_QPE}), there remains the question as to whether estimation can really be well-defined in such situations.

This feeling speaks to a certain view of estimation -- which we have labeled `conservative' (cf.\ Sec.~\ref{sec:conservative_pragmatic}) -- according to which \textit{there is something about the very nature of estimation} that precludes applying it in situations where the estimand is being modeled as a quantum variable and not a parameter. Whereas the pragmatist says `if we \textit{can} do it consistently, why shouldn't we?', the conservative replies, `because it is against the very spirit of estimation!'

Moreover, the conservative should not be satisfied if Entanglement were avoided by making an \textit{ad-hoc} restriction to situations in which the quantity can be viewed unambiguously as a parameter. If the scope of estimation is to be so restricted, then this should be done on the basis of some general principle, not in a reactionary way -- by burying one's head in the sand, as it were. This raises the question: is there any general principle that one could posit as a defining feature of estimation that excludes modeling the estimand as a quantum variable from the outset?

In this section, we identify an assumption (actually a composite of two more basic assumptions) that we propose captures what a conservative view could mean by `the spirit of estimation' that we gestured towards earlier. We argue that this assumption is \textit{implicitly} bundled together with the common understanding of what `estimation' means. Once this extra assumption is taken into account, we will see that it actually forces one to \textit{reject} Entanglement. On one hand, this allows us to ground the conservative view in a positive statement about what estimation should mean; on the other hand it throws into sharp relief what must be sacrificed if one wishes to adopt the more permissive pragmatic stance.

Let us reconsider the characterization we gave of estimation in Sec.~\ref{sec:c_control} as a 3-stage protocol consisting of Preparation, Imprinting and Measurement stages. While this presentation is completely standard, we wish to draw attention to two further aspects that are normally left \textit{implicit} in this characterization. 

First, the probe systems are prepared in an identical manner and interact with the system $A$ in the same way. This suggests that we are \tit{indifferent} with regards to the labeling of the probes, that is, given a set of measurements on the probes $X_1,\dots,X_N$, the results of the estimation should not be altered if one were to arbitrarily permute the ordering of the probes before measuring them. Formally, this implies that the quantum state of the $N$ probes after imprinting is implicitly assumed to be \tit{permutation invariant} for all $N$.

Second, note that the number of probes $N$ does not need to be fixed in advance. That is, suppose we had carried out estimation using $N$ probes, and subsequently decided that we wished to improve our precision by extending the estimation protocol to include a further $M$ probe systems. It is implicitly assumed that we can simply continue the protocol, that is, we need only prepare, imprint, and measure a further $M$ probes: we do \tit{not} need to start the whole experiment again from scratch with a fresh preparation of $N+M$ probes. Formally, this implies that the quantum state of the $N$ probes after imprinting can always be obtained from a larger state of $N+M$ probes by taking the partial trace over the last $M$ subsystems of the latter; moreover this should hold for all values $N,M$. 

The latter condition, as a formal requirement, can trivially be satisfied if no other conditions are imposed; for given an arbitrary state $\rho^{(N)}$ one can always extend it to $\rho^{(N)} \otimes \rho^{(M)}$, where the state $\rho^{(M)}$ is completely unconstrained, and this will evidently reduce to the original state when the appended state is `traced out'. However, it is easy to see that an extension of this form in general cannot satisfy the \tit{permutation invariance} requirement mentioned above. (For instance, the trivial extension just mentioned implies that the sets of $N$ and $M$ probes are statistically independent; it follows that permutations that exchange probes between these two sets will spoil any correlations that existed among the probes within each set.) 

The \tit{joint} assumptions of permutation invariance and extendibility to arbitrary $N$ therefore imply a quite strong condition on the possible quantum states of the probes after imprinting. 

We may formalize these notions as follows. Consider the joint state $\rho_{AB}$ of some system $A$, with a set of $N$ systems $B$, such that the systems $B$ are invariant under permutations. Call this state $M$\tit{-extendable} if it can be further extended to include another $M$ systems, such that the state of the $N+M$ systems (that is, excluding system $A$) are also invariant under permutations. 

Note that the state of the probes $\rho^{\trm{out}}$ as given in Eq.~\ref{eqn:probe_out} is readily seen to be permutation invariant, and if it is taken alone, i.e.\ not considering $A$, it is also $M$\tit{-extendable} for all integers $M$. However, this does not guarantee that these properties automatically continue to hold when one includes the system $A$ in one's considerations. We suggest that in most examples of quantum parameter estimation, it is \tit{tacitly assumed} that they do --- that is, that the following methodological assumption is adopted: \\

\tbf{Extendibility:} The joint state of the register $A$ and probes $B$ after the imprinting step is $M$\tit{-extendable} for all integers $M$.\\

We may now state our key result of this section: according to Theorem 1 of Ref.~\cite{doherty_2004}, if Extendibility holds, then the joint state $\rho_{AB}$ of the system $A$ and the $N$ probes $B$ must be a \tit{separable} state: it cannot be entangled. Consequently, this assumption excludes Entanglement.

The result just cited has the following consequence for our argument. On the one hand, within a general quantum model like the one we have proposed, Entanglement is generically expected to hold, as we argued in Sec.~\ref{sec:implications}. On the other hand, it seems natural to take Extendibility \textit{as part of the definition of estimation}. That is, one may wish to assert that the 3-stage estimation protocol is only well-defined against the background condition that (i) one is indifferent to the ordering/labeling of the probes; and (ii) one is indifferent as to the precise number $N$ of probes that will be used. In that case, estimation is conceived as presupposing Extendability from the start, and even a quantum generalization of it must respect this principle and so should exclude Entanglement. 

(\textit{Remark:} Here we must address a possible point of confusion. There is a situation where it might seem that estimation is flatly incompatible with Extendibility: when the probes are permitted to be entangled. As an example, $N$-partite states of the ``GHZ" type $\ket{\Psi^{(N)}} := \frac{\ket{0}^{\otimes N} + \ket{1}^{\otimes N}}{\sqrt{2}}$ are known to be useful for estimation, yet this state is blatantly non-Extendible: adding more probes renders it non-invariant under permutations. We note, however, that in theoretical analyses one usually lumps together the entangled probes into $N$ independent `blocks' of $M$ entangled probes, resulting in a state of the form: $\ket{\Psi^{(M)}}^{\otimes N}$. This can be regarded as Extendible under a minor adjustment of terminology: just let `probe' refer to each $M$-partite entangled block, rather than to its constituents. This may be seen as a refinement of the definition of `probe' that goes along with choosing to take Extendibility as a background condition of estimation.)

In support of this idea, let us consider how peculiar estimation looks when these assumptions are given up --- as they must be, in a scenario where Entanglement is deemed to hold. First, it means that expectations which apply when there are less than some threshhold number $N'$ of probes may no longer be valid when one exceeds this special number; in particular one may find that even if one is initially indifferent to the probe ordering, it suddenly becomes important when the number of probes exceeds $N'$. If nothing else, this implies that one's model of the phenomena must have a great deal more structure than the simple 3-stage paradigm would lead one to suspect: the model must indicate a specific threshhold $N'$ and prescribe what happens under permutations when invariance no longer applies. All of this will be encoded, naturally, in whatever joint quantum state one assigns to $A$ and $B$. This reveals the beauty and efficiency of Extendibility: when justified, it cuts through all this complexity by radically restricting this joint state to a highly symmetric form.

By explicitly recognizing Extendibility as an additional component of the definition of estimation, one can consistently reject Entanglement, justifying the conservative approach that limits the scope of ``estimation".

However, this move can only be consistent if it is \textit{explicit}, for then it can be used as a test for whether a given scenario meets the conditions required for this stronger notion of estimation. Otherwise, one runs the risk of tacitly assuming Extendibility holds --- thus that Entanglement is false --- even in scenarios where a more careful analysis would reveal that the probes and the relevant degrees of freedom of the system $A$ that carry information about $\theta$ should in fact be modeled as entangled. In that case, we have shown there is a manifest inconsistency with the assumption that estimation reveals a unique and independent pre-existing value; this inconsistency may escape notice if one is operating on the tacit (unjustified) assumption that Extendibility holds.

Finally, in experimental scenarios where evidence of this inconsistency might become measurable --- for instance, where some degree of experimental access to the system $A$ becomes feasible, as in the case of the field produced by superconducting qubits in superposition --- one may use the criterion of Extendibility as a statistical test to check whether, in fact, the inconsistency obtains. For if the joint statistics of measurements on $A$ and $B$ satisfies Extendibility, we know that Entanglement can be safely rejected, allowing one to consistently posit a unique unknown value that exists before and independently of measurement.

\subsection{Implications of rejecting the assumptions}\label{sec:implications}

We have argued that a set of assumptions natural to quantum parameter estimation---Uniqueness, Pre-existing Value, Independence, Entanglement, and Completeness---cannot all be jointly sustained. We now consider whether one or more of these assumptions might be rejected to avoid the contradiction. 

\textit{Reject only Completeness}. One way to avoid the contradiction is to deny that the quantum state contains complete information about the outcome of a sufficiently accurate measurement that could be performed. As noted in Sec.\ref{sec:problem_QPE}, rejecting Completeness while maintaining the other assumptions implies that $\theta$ constitutes a hidden variable for the initial state of system $A$. While a variety of consistent hidden-variable accounts of quantum theory do exist, they all have controversial features (for a review, see eg.\ Sec.8 of~Ref.\cite{sep-bell-theorem}). In light of the no-go theorems, one would have to carefully weigh the benefits of retaining a realist view of estimation against the potential costs of giving up Completeness, such as non-locality. This approach is likely to appeal only to those who already have independent reasons to prefer a hidden-variable interpretation.

\textit{Reject only Entanglement.} 
We observed in Sec.~\ref{sec:problem_QPE} that Entanglement follows from the more basic `background' assumption that estimation should be defined in terms of the circuit model depicted in Fig.~\ref{fig:PE_circuit_qm}. To reject Entanglement, it is necessary to replace this definition of estimation with a more restrictive one, for example, by including an Extendibility assumption as we proposed in Sec.~\ref{sec:de_finetti}.

The main disadvantage of this approach is that, while it secures the consistency of the realist view of estimation, it does so essentially by \textit{fiat}: if in a given situation there is even the slightest doubt about whether the quantity of interest can be modeled as a parameter, one must give up estimating it. In light of our main theorem, this seems to constrain the scope of estimation unnecessarily. It means, for instance, that the question we posed at the beginning -- what can we do if we are \tit{uncertain} about whether to model the quantity of interest as a parameter or a quantum variable? -- is impossible to answer within the scope of estimation. 

Yet, our no-go theorem actually shows that one \tit{can} make consistent and meaningful claims in such a situation, provided one is explicit about which of the remaining assumptions one chooses to give up. This decision will constrain the kinds of knowledge claims one can consistently make about the estimand -- but it does not make estimation itself impossible. Rejecting Entanglement therefore misses an opportunity to say something constructive about estimation in the quantum setting.

\textit{Reject the realist commitment to estimation}. Against the realist intuition that motivates the assumptions of Uniqueness, Pre-Existing Value, and Independence, one might argue that $\theta$ \tit{only} functions as a parameter indexing a family of statistical models used for inference, and \tit{never} does `double duty' as a fact constraining the real properties of a system independently of measurement. This is tantamount to a shift in how one regards measurement generally: instead of the broadly realist view that sees measurements as informing us about \tit{how things are}, one takes a broadly coherentist view that measurements primarily inform us about the likely responses \tit{of other measurements}, as captured in the statistical model one is using -- any inferences from the statistical model to an underlying reality must be made cautiously and are only justified insofar as they are pragmatically useful. 

This view has strong precedent in foundational texts. For instance, in \textit{Probabilistic and Statistical Aspects of Quantum Theory} \cite{holevo2010probabilistic} Holevo explicitly cautions against reading $\theta$ as a physically real quantity\footnote{``It is worthwhile to emphasize here again that statistical model is a mathematical object. [...] [I]n general there is no guarantee that any quantum-theoretic state or measurement can be physically implemented, even in principle. [...] The relevance and the usefulness of quantum theory are ultimately conrmed by the continuing success of its application [...] any general result obtained in the framework of quantum theory wittingly applies to “implementable” states and measurements in so far as quantum theory gives a correct model for reality.'' --Holevo,~\cite{holevo2010probabilistic} }. He treats it instead as a statistical index, not assumed to have any meaningful determinate value independently of measurement. 

According to this view, one gives up on the realist commitment regarding the results of estimation, so that the assumptions of Uniqueness, Pre-existing Value, and Independence no longer necessarily apply. Such a view is also readily accommodated by interpretations of quantum theory that follow in the spirit of the Copenhagen school (associated with the writings of Bohr, Heisenberg, Pauli and others). In particular, contemporary ``Copenhagenish"~\cite{schmid_copenhagenish_2025} interpretations explicitly regard quantum states (and, by extension, parameterizations of them) as strictly model-based inferential objects whose primary function is to inform predictions about the outcomes of future measurements.

This strategy dissolves the contradiction, but at a substantial conceptual cost. Precision measurement in metrology is widely understood to aim at tracking real features of physical systems—not merely updating beliefs within a formal scheme. For instance, a landmark paper on quantum metrology in the presence of noise asserts that the \tit{error} of a parameter $x$ is a function of the average squared difference $(x_{\trm{est}}-x_{\trm{real}})^2$, where (according to the authors) ``$x_{\trm{real}}$ is the real value of the unknown parameter $x$ and $x_{\trm{est}}$ is the estimated value of $x$"~\cite{escher_general_2011}. We note that if $x_{\trm{real}}$ fails to be either Unique, Pre-Existing, or Independent, this definition makes little sense. We suspect that many physicists implicitly endorse a coherentist view -- a recent survey shows the Copenhagen interpretation is still widely endorsed~\cite{jedlicka_survey} -- in which case our result should be taken as a caution against uncritically carrying over the ``true-value" concept from classical statistics to the quantum domain.

\textit{Reject only Uniqueness.} We have formulated Uniqueness with definitional uncertainty in mind (see Section~\ref{sec:problem_QPE}). Therefore, merely allowing that a sufficiently accurate measurement be consistent with a range of possible values of $\theta$ does not constitute a rejection of Uniqueness. Rather, such a rejection entails something much stronger: two hypotheses that could in principle be discriminated by a sufficiently accurate measurement, call them $\theta'$ and $\theta''$, must nevertheless be simultaneously true. This creates a profound puzzle, for in the limit of large $N$, the estimation is expected to converge to a unique value of $\theta$: should it then be $\theta'$ or $\theta''$? And how can the other hypothesis remain true, given that it is not consistent with the results actually obtained?

One view that may accommodate this conclusion is the \tit{many-worlds interpretation} of quantum theory~\cite{Vaidman_MWI}. On this view, a measurement does not result in a unique outcome: rather, as many ``worlds'' exist (or come into existence, depending on the variant) as there are possible results of a measurement, and in each world is a copy of the experimenter who observes that particular result. Thus, while we may obtain measurement data in the limit of large $N$ that converges to $\theta'$, elsewhere in a parallel sector of the universe inaccessible to us, our clones have obtained data that converged instead to $\theta''$. Both hypotheses are true, but we only see one. This interpretation has its defenders, but also faces numerous criticisms in the literature, not least that it may strike many readers as metaphysically extravagant~\cite{Kent_2010}.

\textit{Reject only Pre-Existing Value.} This possibility is interesting because Pre-Existing Value can be decomposed into an `if' and an `only if' clause, and different metaphysical commitments may lead one to reject one and not the other of the two. We discuss these in turn. (Note that in both cases we are still assuming Uniqueness holds at the time of measurement, though it need not hold at earlier times):

\tit{Option (i):} Allow that a hypothesis $\theta \in [\theta']$ is true at the time of measurement \tit{only if} it was true before, but not conversely. 

In this case, we allow that multiple hypotheses may be true prior to measurement, but that the measurement act selects one as true and makes the rest false. This fits well with ``dynamical collapse'' theories: modifications of quantum theory that interpret the superposed quantum state as representing multiple true values of $\theta$ prior to measurement, and which introduce an ad-hoc mechanism that collapses the state dynamically when certain conditions are met that are typically associated with performing a measurement, so that only one value remains true when the measurement is complete (for a recent review of such proposals, see Ref.~\cite{Bassi_collapse}). 

While they have a certain appeal, objective collapse theories suffer well-known problems. One is that they make predictions that deviate from those of standard quantum theory, since they introduce a physical mechanism of ``dynamical collapse'' not accounted for by the latter. Within the regimes that have been tested so far, no objective collapses have been detected. Moreover, theoretically, the non-linear character of the collapse mechanism can create problems for the probability interpretation of such theories, and may introduce physical non-locality~\cite{gao_2023}. This option is therefore somewhat radical; one would hope that a consistent account of quantum estimation could be given without needing to go beyond quantum theory.

\tit{Option (ii):} A hypothesis $\theta \in [\theta']$ is true at the time of measurement \tit{if} it was true before, but not conversely. On this account, measurement can make a hypothesis true that was previously false. Either some other hypothesis was previously true and then the measurement changed it (`disturbance'), or \tit{no} hypotheses were true, and the measurement contributed somehow to generating the true hypothesis (`creation'). Interpretations of the former kind are usually hidden variable theories, in which the ability of a measurement to change the value of a hidden variable is essential to explaining how they can reproduce the predictions of quantum theory -- however they are for this reason non-local, since measurements on one member of an entangled pair can instantaneously disturb the hidden variables at the site of its partner. The `creation' option potentially fits with some modern interpretations inspired by the ``Copenhagen'' school, if one reads them as stating that a quantity in superposition simply \tit{has no determinate value}, unique or otherwise, until it is measured, whereupon it spontaneously acquires a unique value (for a recent review, see Ref.\cite{schmid_copenhagenish_2025}). 
 
\textit{Reject only Independence}. Rejecting Independence while \tit{maintaining} the other assumptions leads to the rather peculiar combination of asserting that, on the one hand, there \tit{is} a unique, true hypothesis $\theta \in [\theta']$ just prior to the imprinting phase, and on the other hand, no conceivable measurement exists that would reliably reveal it --- not even a highly idealized `sufficiently accurate' measurement whose precision is limited only by definitional uncertainty. 

While not inconsistent in a strict logical sense, this position appears to make the very idea of measurement \tit{incoherent}: it breaks the role of measurement as a truth-preserving inference procedure. It appears to us quite meaningless to assert that $\theta \in [\theta']$ is true, if there is no hypothetical measurement that could reliably single out $\theta'$ from among the alternatives in $\Theta$.

\section{Conclusions \label{sec:conclusion}}

Quantum parameter estimation is often presented as an interpretation-neutral framework for extracting information about physical quantities. Our results suggest that the superficial appearance of neutrality is misleading: the conceptual foundations of quantum parameter estimation, once made explicit, are intimately bound up with the foundations of quantum theory, and particularly the interpretation of quantum states. 

We drew attention to the common framing of estimation as the discovery of an unknown value -- as though it were uniquely fixed prior to and independently of the estimation process -- making it explicit via the assumptions of Uniqueness, Pre-existing Value, and Independence, whose conjunction we argued forms the basis for a `realist commitment' to estimation. We then demonstrated that if one broadens the definition of estimation to include estimation of physical quantities that may be represented as quantum variables in a more general experimental setting, then the realist commitment is in conflict with certain assumptions about the meaning and application of the quantum formalism, as captured by the Entanglement and Completeness assumptions.

Our analysis addresses a basic question: what precisely should we take to be the conceptual foundations of parameter estimation?  We showed how consistency can be restored by rejecting only \tit{one} of the full set of five assumptions, and how the decision of which assumption to reject is directly related to the question of how quantum states relate to reality, that is, the very same question that animates debates in quantum foundations. 

Our result also clarifies what is really at stake in choosing between a `conservative' versus a `pragmatic' understanding of quantum estimation. On the one hand, if we take the pragmatic view that estimation is nothing more than a three-stage experimental protocol in which measurements on the probe systems are used to make a conditional evaluation of a physical quantity, then there is no reason why the estimated quantity cannot itself be modeled as a quantum variable: we demonstrated how to do this in an explicit model. This allows us to extend the scope of estimation considerably, but at the cost of requiring that we take an explicit stance regarding the interpretation of the quantum formalism, to avoid inconcistencies.

On the other hand, we showed how one could sidestep thorny interpretational problems by adopting a `conservative' view, based on adopting a strengthened definition of estimation that entails an explicit assumption of Extendibility. Since this assumption is proven to exclude Entanglement, the contradiction identified by our no-go theorem can be avoided, at the price of restricting the scope of estimation to only those quantities and experimental conditions that meet the stringent requirements of the Extendibility assumption. 

While we do not argue for the superiority of one approach over the other, our analysis brings to light the relevant assumptions and consequences of each approach, enabling researchers to make an informed decision about how to frame quantum estimation both as an operational procedure \tit{and} as an epistemological endeavor, while steering clear of pitfalls and inconsistencies. Perhaps the most salient message for most quantum estimation researchers is that one should be wary of defining estimation errors in terms of difference from a hypothetical ``true value", if one does not wish to commit oneself to a realist commitment regarding the estimand. Viable alternatives exist, for instance, there is a precedent for framing estimation in the language of hypothesis testing, which leaves open what the hypotheses may be about (thus permitting agnosticism about the connection between models and reality). As for the definition of estimation error, this can be replaced with a Bayesian notion of variance in the posterior distribution over possible values, avoiding the need to invoke a notion of ``true value"; this is indeed the approach adopted in the most recent edition of the Guide to Uncertainty in Metrology (GUM)~\cite{JCGM_GUM_2008}. 

This strategy may appeal to those who wish to focus on operational aspects of estimation, such as optimizing the precision of an estimate, as it allows one to defer questions about the meaning of the estimated value \tit{vis-a-vis} real objective facts about the system. Provided one is only interested in predicting the outome statistics for various estimation protocols, one then need not trouble oneself with matters of epistemology. 

In contrast, our message for researchers in quantum foundations is that quantum estimation is a fertile ground for new results and debates, precisely in case one \tit{is} interested in how the estimated value informs us about an independent reality. As a case in point, our work shows that researchers who wish to endorse a `realist commitment' (as we have defined it) must either adopt a `conservative' definition of estimation founded on the Extendability assumption (Sec.~\ref{sec:de_finetti}), or else conclude that the quantum state is incomplete -- widely regarded as a niche (if not radical) position. Giving up on the realist commitment is no less demanding, since as we have shown, it entails quite non-trivial assertions about the meaning of quantum states, depending on which `realist' assumption one chooses to abandon.

In conclusion, we expect that our results and the conceptual framework in which they were derived will provide the basis for a new dialog between the disparate subfields of quantum metrology and quantum foundations. This is fitting, since both are, in their own distinct ways, concerned with the same ``measurement problem".

\begin{acknowledgments}
This publication was made possible through the support of Grant 62424 from the John Templeton Foundation. JP acknowledges support from the Conselho Nacional de Desenvolvimento Científico e Tecnológico (CNPq) Grant/Process number PV351823/2025-5. 
\end{acknowledgments}


\end{document}